\documentclass[prb,aps,amsfonts,amssymb,floatfix,showpacs,superscriptaddress]{revtex4}
\usepackage{graphicx}
\usepackage{dcolumn}
\usepackage{bm}
\usepackage{color} 
\usepackage{epsfig}
\usepackage{epstopdf}
\begin{document} 
\title{
Excitonic insulators as a model of $d-d$ and Mott transitions in strongly correlated materials}
\author{R.S. Markiewicz and A. Bansil}
\affiliation{ Physics Department, Northeastern University, Boston MA 02115, USA}
\begin{abstract}
We show how strongly correlated materials could be described within the framework of an excitonic insulator formalism, and delineate the relationship between inter- and intra-band ordering phenomena. Our microscopic model of excitons clarifies the fundamental role of Van-Hove-singularity-nesting in driving both inter- and intra-band ordering transitions, and uncovers an interesting connection with resonating-valence-bond physics. 
\end{abstract} 
\maketitle

\section{Introduction}

At the mean-field level, density-wave instabilities and excitonic insulators are generally understood to be two different faces of the same phenomenon.  For example, both ferro- and anti-ferromagnets can be considered as materials, which have undergone magnon condensation.  It is important to understand how these considerations play out when we go beyond the mean field level. For example, in the strong-coupling limit, are there always preformed excitons, analogous to Cooper pairs in the Bose-Einstein condensation (BEC) limit of a superconductor?  Is the transition always driven by a soft mode?  Are there differences between the conventional picture of excitonic insulators in a two-band semiconductor and a one-band system?  After all, charge- and spin-density waves are common to both limits.

Strongly correlated materials could be defined as materials, which have an anomalously small ratio of $k_BT_c/2\Delta$. In a study of charge-density-wave formation, McMillan\cite{McMi} showed that this anomalous ratio can be associated with phonon entropy.  As many phonons try to soften at the same time, they interfere with one another, leading to an extended temperature range where only short-range order is present.  This effect is now calculated via a mode coupling theory\cite{Moti}.  Using Moriya's electronic version of mode-coupling theory\cite{Moriya}, a similar effect is found to account for the pseudogap in cuprates\cite{MBMB}, except that here the entropy arises from competing {\it electronic} bosons -- fluctuating spin-or charge-density waves.  In cuprates, this large entropy is associated with a broad susceptibility plateau near $(\pi,\pi)$.  Since delocalization in momentum space implies localization in real space, the question arises whether the pseudogap represents a form of exciton condensation.  Exciton condensation\cite{Mott,KeKu,Kohn,HaRi,CoG,BroF} was introduced in the 1960's as a model of strong correlation in solids, wherein excitons are present in the ground state.  However, to cast the problem in hydrogenic form the bands were generally assumed to be parabolic, so that the interaction $V$ was taken to be small compared to the bandwidth, and real band effects were neglected.

Excitons are typically associated with electrons and holes in different bands, whereas the cuprates are dominated by a single band.  To better understand single-band excitonic insulators, we analyze a two-band analog of the cuprates.  Noting that orbital ordering can be described by a pseudospin model, we study the problem of excitonic instability between two $d$-bands.  We find that when realistic band structures are assumed, the properties of excitonic insulators are significantly modified.  In particular, the excitonic hybridization leads to strong mixing of the bands in the vicinity of their respective Van Hove singularities (VHSs), leading to an orbital order.

There is a well-known analogy between exciton condensation and superconductivity, namely, the crossover from weak-coupling (BCS) superconductivity to a strong-coupling form involving preformed Cooper pairs and subsequent Bose-Einstein condensation (BEC).  In the excitonic case, the semiconductor phase corresponds to BEC, where an instability occurs when the excitonic binding energy $E_X$ is larger than the indirect gap $E_g'$ between the filled lower (valence) band and the empty upper (conduction) band, leading to a phase of preformed excitonic pairs.  For $E_g'<0$, the material becomes a semimetal with a weaker, partially screened excitonic phase corresponding to BCS.  We find that the semimetal phase is absent in the undoped case when realistic bands are assumed, replaced by a striking anticrossing phenomenon.  Remarkably, in the one-band analog\cite{MBMB}, we find that a BEC-BCS transition, akin to a  Mott-Slater transition, can be driven by tuning the band structure.

More recently, interest in excitons has been revived both by improved calculations of the Bethe-Salpeter equation and the great progress in making monolayer materials.  In particular, $E_X$ is enhanced in 2D materials, and the issue of strong correlations in graphene has hinged on whether or not graphene has an excitonic insulator (EI) ground state.  Kohn argued that the EI could explain Mott transitions\cite{Kohn2}, and for a while Mott seemed convinced.\cite{Mott2}   Our analysis here confirms their insight with realistic band-structure calculations.

\section{Results}

\subsection{$d-d$ Excitonic insulators}

We start with a model relevant to many correlated materials -- an excitonic transition between two bands of predominantly $d$-character, in particular the $e_{2g}$ bands $d_{x^2-y^2}$ and $d_{z^2}$.   Our basic Hamiltonian is $H=H_0+H_{e-e}$, with
\begin{equation}
H_0=\sum_{{\bf k},i=v,c}\epsilon_i({\bf k})a^{\dagger}_{i,{\bf k}}a_{i,{\bf k}},
\label{eq:1}
\end{equation}
with $i =v,c$ for the valence (v) and conduction (c) band, and
\begin{equation}
H_{e-e}=\frac{1}{2}\sum_{\bf l_1,l_2,l_3,l_4}V_{\bf l_3,l_4}^{\bf l_1,l_2}a^{\dagger}_{\bf l_1}a^{\dagger}_{\bf l_2}a_{\bf l_4}a_{\bf l_3},
\label{eq:2}
\end{equation}
with ${\bf l}=i,{\bf k}$.  We use a Thomas-Fermi screened potential $V_q=2e^2/\epsilon_0a_0\sqrt{(q^2+q_s^2)}$ with $\epsilon_0$ the background dielectric constant, $a_0$ the in-plane lattice constant, and choose a weak screening wave vector $q_s=1.2\pi/a_0$ to avoid a $q=0$ divergence.  For simplicity, we neglect the exchange energy, which leads to a splitting in energy of singlet and triplet solutions, so that electron spin need not be explicitly considered, and we restrict the calculations to half filling where screening is weakest.

For ease in comparing with one-band cuprate results, we first assume that both bands have the same dispersion, using hopping parameters which fit the dispersion of La$_{2-x}$Sr$_x$CuO$_4$ (LSCO),  
\begin{equation}
\epsilon_i({\bf k})=-2t(c_x(a)+c_y(a))-4t'c_x(a)c_y(a)-2t''(c_x(2a)+c_y(2a))-2t'''(c_x(3a)+c_y(3a)),
\label{eq:1b}
\end{equation}
with $c_i(\alpha a)=\cos(k_i\alpha a)$, $i=x,y$, and $\alpha$ is an integer, and $t=0.21$~eV, $t'/t=-0.089$, $t''/t=0.043$, and $t'''/t=0.081$.  In this case, the valence band has a single maximum at $Q=(\pi,\pi)$ and the conduction band a single minimum at $\Gamma$ with a direct gap $E_g$ between them.  The model has one anomalous property, that for $q=0$, the joint density of states (JDOS) $\sim\delta (E-E_g)$.  However, since it is intended to reproduce the conventional excitonic nesting at $Q$, this should not be a problem.   The gap equation becomes
\begin{equation}
\Delta_k=\int\frac{dk'_xdk'_y}{4\pi^2}V_{k-k'}\chi_{Q,k'}\Delta_{k'},
\label{eq:3}
\end{equation}
where 
\begin{equation}
\chi_{Q,k}=-\frac{f(E_{+})-f(E_-)}{E_{+}-E_-},
\label{eq:4}
\end{equation}
with $f(E)$ the Fermi function, $E_{\pm}=\epsilon_+\pm\sqrt{\epsilon_-^2+\Delta_k^2}$, and $\epsilon_{\pm}=(\epsilon_v({\bf k})\pm\epsilon_c({\bf k+Q}))/2$,
and $\mu$ is adjusted to keep the same density as for our starting point of full valence band and empty conduction band.  
To simplify, we assume that $\Delta_k$ is isotropic, in which case Eq.~\ref{eq:3} becomes
\begin{equation}
1=<\int\frac{dk'_xdk'_y}{4\pi^2}V_{k-k'}\chi_{Q,k'}>_k.
\label{eq:5}
\end{equation}
We have confirmed that the integrand $V\chi$ is only weakly $k$-dependent.  

\begin{figure}
\leavevmode
\epsfig{file=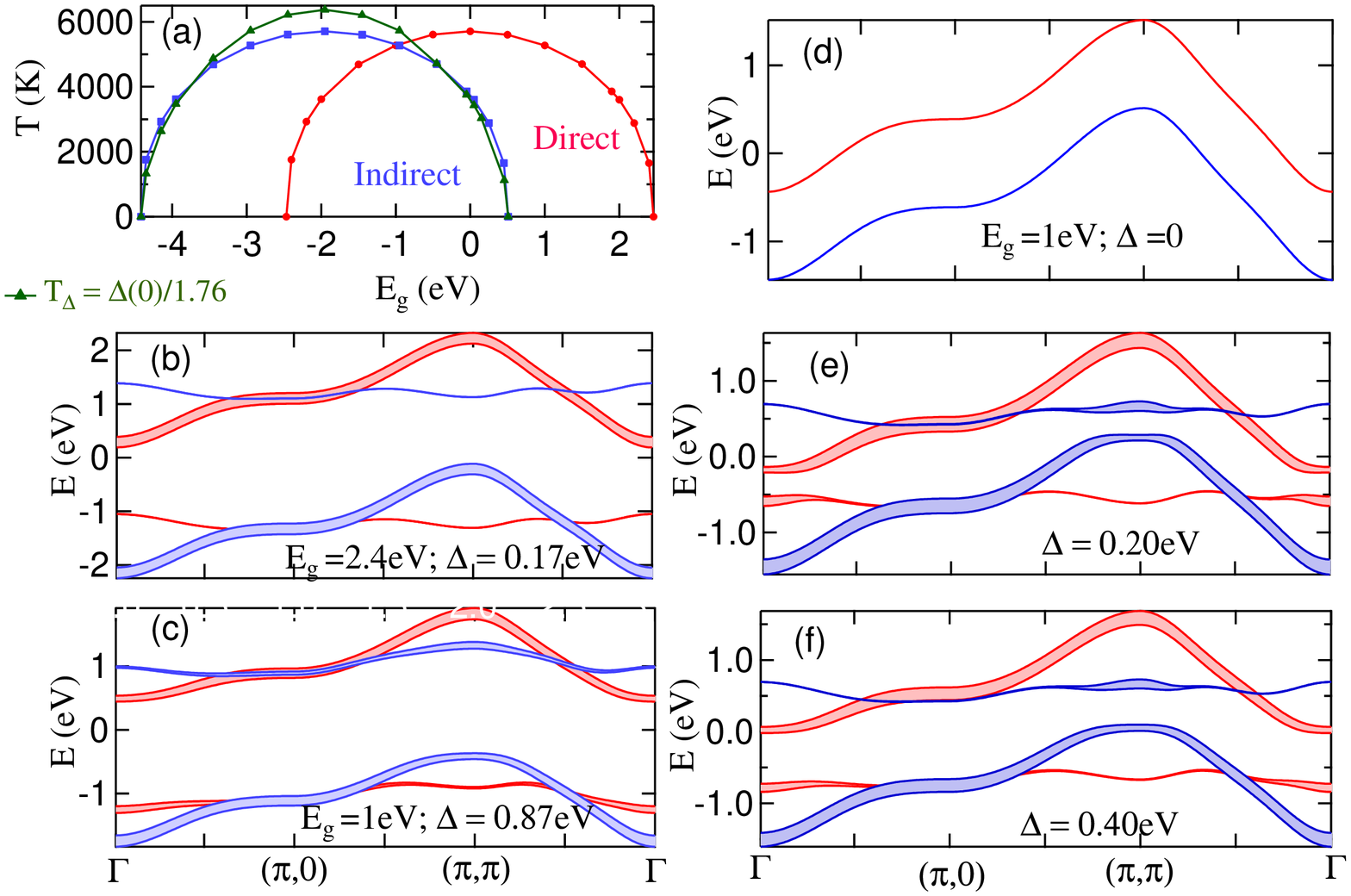,width=11.5cm, angle=0}
\vskip0.5cm  
\caption{
{\bf Excitonic insulator crossover for two bands of same dispersion,} Eq.~\ref{eq:1b}.
(a) Plot of $T_c$ (red line with filled circles and blue line with squares) and $\Delta$ (green line with triangles) vs the indirect gap $E'_g$ (blue and green lines) or the direct gap $E_g$ (red line). (b,c) Corresponding dispersions at $T=0$ for $E_g$ = 2.4~eV, $\Delta(0)$ = 0.17 (b) and $E_g$ = 1.0~eV, $\Delta(0)$ = 0.87~eV (c).
(d-f)  Continuation of dispersion (c) to higher temperatures, $T$ = 5200 (d), 4900 (e), and 4400K (f).  In frames (b-f), the color indicates features derived from the conduction (blue) or valence band (red), while widths of lines indicate relative spectral weights.
}
\label{fig:2}
\end{figure}
\begin{figure}
\leavevmode
\epsfig{file=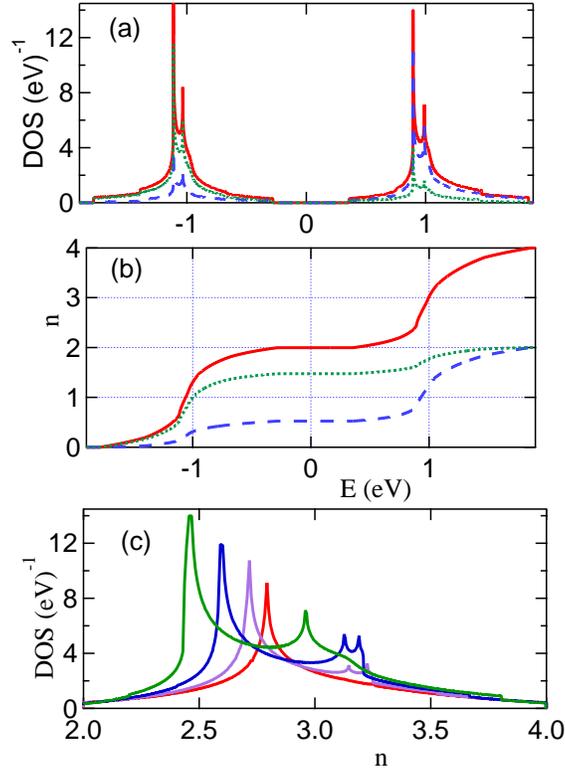,width=8.5cm, angle=0}
\vskip0.5cm  
\caption{
{\bf Density and DOS for excitonic insulator phase of Fig.~\ref{fig:2}.}
(a,b) Plot of DOS (a) and density (b) vs energy for the data of Fig.~\ref{fig:2}(c) showing contributions due to the lower bands (blue dashed lines) and upper bands (green dotted lines) and their sum (red solid lines).  (c) DOS of VHS peaks of upper bands, for data corresponding to Fig.~\ref{fig:2}, frame (d) (red), (e) (violet), (f) (blue), and (c) (green).  
}
\label{fig:3}
\end{figure}

The resulting phase diagram is shown in Fig.~\ref{fig:2}(a), as $T_{\Delta}=2\Delta(T=0)/3.53$ and $T_c$ plotted against the direct gap $E_g$ or the indirect gap $E'_g$.  We note that $2\Delta(0)/k_BT_c$ has approximately the BCS ratio, which is not always the case when the two bands have different dispersions.   Figures~\ref{fig:2}(b,c) show the resulting $T=0$ dispersion at several values of $E_g$, revealing a striking difference from the electron gas result.  For all values of $E_g$, the resulting excitonic gaps are nearly identical, and the ground state is always fully gapped (insulating). At finite temperatures, Figs.~\ref{fig:2}(d-f), the gap shrinks, and a more conventional semimetallic behavior is restored.  One can understand what is happening by looking at the density of states (DOS), Fig.~\ref{fig:3}, where we see that the excitonic state is characterized by splitting of the VHSs of both bands.  We have tested this in a number of different model dispersions, and typically find very similar results.  This is a plausible result, since the VHSs are typically the strongest excitonic features observed in optical spectra.\cite{JCP,Rief}

We note that as $E_g$ changes, the excitonic gap is a symmetric function about its peak value, Fig.~\ref{fig:2}(a).  Indeed, the {\it dispersions}, Figs.~\ref{fig:2}(b,c) are also symmetrical, and the bands never cross.  The excitonic order parameter is $\sim <a_{c,k+Q}^{\dagger}a_{v,k}>$, representing hybridization of the conduction and valence bands.  This hybridization leads to an anticrossing phenomenon: as $E_g$ varies, the two bands sit at a fixed energy separation, and the valence and conduction bands gradually interchange their orbital character. Notably, we find that the excitonic gap is largest when $E_g=0$ and the two bands would perfectly overlap in the absence of Coulomb interaction.  This suggests a close connection to the corresponding theory of Van Hove nesting of intraband excitons, where the two bands are the spin up and spin down bands, and hence necessarily degenerate.  This will be discussed further below.

\begin{figure}
\leavevmode
\epsfig{file=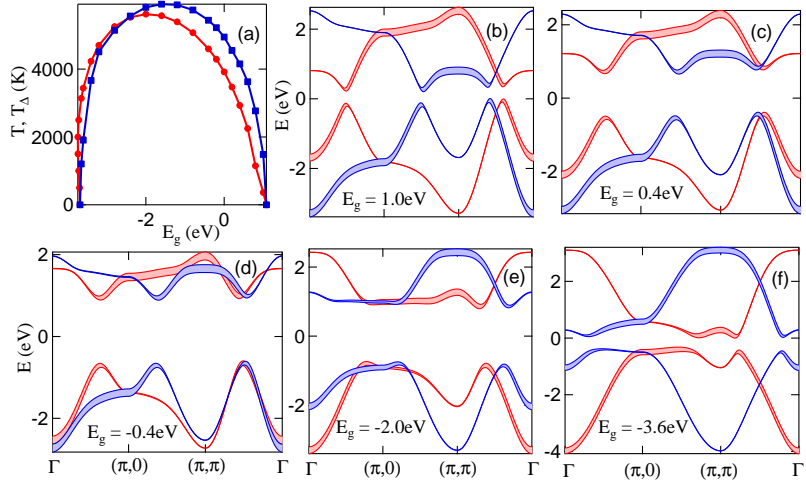,width=11.5cm, angle=0}
\vskip0.5cm  
\caption{
{\bf Excitonic insulator crossover for manganite $e_{2g}$ bands.}
(a) Plot of $T_c$ (red circles) and $T_{\Delta}$ (blue squares) vs gap $E_g$. (b-f) Corresponding dispersions at $T=0$ for several values of $E_g$.
}
\label{fig:4}
\end{figure}

We have repeated the calculation for a model of the coupled $e_{2g}$ bands of a layered manganite\cite{Baubl}, Fig.~\ref{fig:4} [Appendix].  A similar evolution dominated by anticrossing phenomena is found, but since the $d_{x^2-y^2}$ and $d_{z^2}$ bands have different dispersions, the gap and $T_c$ do not exactly follow the BCS ratio.  Indeed, the two curves peak at different values of $E_g$, and $T_c$ actually shows reentrance phenomena on the lower side of its dome.  Despite these differences in detail, the gross features of the two models are very similar.
The conventional excitonic insulator results are recovered if $V$ is small compared to the bandwidth, and the VHSs are too energetically distant.  However, the parabolic band approximation makes anticrossing behavior topologically forbidden: an upward-curving parabola cannot turn into a downward-curving parabola. 

These results suggest a number of reasons why excitonic transitions are hard to see.  First, the peak transition temperatures are so high that one would only notice the transition by tuning close to where the gap closes.  We note that the high $T$-scale is characteristic of 2D excitons, which tend to have larger binding energies.  Secondly, one usually looks for signatures of a transition near the Fermi level, but here is an insulator-to-insulator transition, with no states near the Fermi level.  Moreover, the strongest signals of the transition are well away from the Fermi level, where they could be smeared out by strong carrier scattering.  This is particularly a problem near the VHS, where broadening is generally quite large.  Yet another complication is the close relation between excitonic insulators and charge/spin density waves (C/SDWs).  In the weak coupling limit the excitonic state evolves into a CDW (singlet exciton) or SDW (triplet exciton), whereas at strong coupling there are conspicuous effects of excitonic binding.

Despite these obstacles, an excitonic insulator may have been found in a paradigmatic 2D CDW, $2H$-NbSe$_2$.\cite{Paco} Quoting from the abstract, ``in the single layer, the CDW barely affects the Fermi surface of the system, thus ruling out a nesting mechanism as the driving force for the modulation. The CDW stabilizes levels lying around 1.5 eV below the Fermi level within the Se-based valence band but having a substantial Nb-Nb bonding character.''  Moreover, their Fig.~4 clearly reveals the splitting of the VHS in the valence band, although a conduction band gap opens away from the corresponding VHS.  All these features are consistent with the present model.

While the present calculations clarify the role of excitonic insulators in producing interband hybridization, there is one important issue we have not addressed. Density-functional theory (DFT) calculations already capture many important aspects of interband hybridization, including avoided crossings.  Thus, care must be taken to avoid double counting of hybridization effects when we attempt to study high-$T$ exciton formation, which is not properly treated in DFT.  This will be of particular interest in those materials where hybridization plays a large role, including heavy-Fermion materials and topological insulators.

\subsection{Orbital vs spin excitons}

\begin{figure}
\leavevmode
\epsfig{file=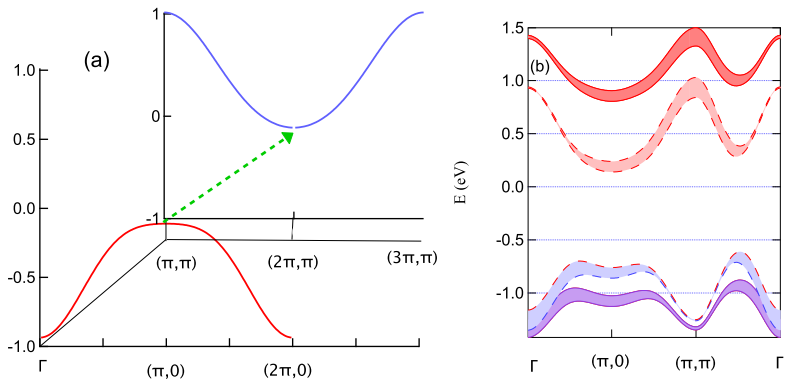,width=11.5cm, angle=0}
\vskip0.5cm  
\caption{
{\bf Intraband excitonic insulator.}  (a) Dispersion across the two VHSs along $X$-direction, showing resemblance to the two-band exciton problem. (b) Dispersions of the same model as in Fig.~\ref{fig:2}, but with $E_g=0$, $\Delta(0)=0.966$~eV (darker shading) or 0.5~eV (lighter shading).
}
\label{fig:12}
\end{figure}

In a recent DFT+MBPT (many-body perturbation theory) study of the cuprate pseudogap\cite{MBMB}, it was found that by adding vertex corrections [excitonic effects] to a GW self energy calculation\cite{AIP} it is possible to account for the competition between density waves (`bosonic entropy' effects) that gives rise to extended domains of only short-range order, consistent with the pseudogap.  The bosonic entropy represents a spread of electron-hole pair weight over many $q$-values, suggesting a localization in real space as excitons.  The density waves fall into two categories, associated with either Fermi-surface nesting or VHS nesting.  Correlation effects are found to be strongest in the presence of VHS nesting, and the competition between the two forms of nesting leads to a sharp crossover between Mott physics and Slater physics -- i.e., from localized to itinerant spins. The high temperature of onset and the lack of connection to the Fermi surface suggest that the commensurate, VHS-related transition is mainly bosonic in character, and we suggested that if the calculation were extended beyond mean-field it would involve preformed excitonic pairs\cite{VHex1,VHex2}.  As $T$ decreases, the VHS nesting is weakened by the Fermi function (Pauli blocking), leading to a crossover to more coherent fermions.  A similar crossover to low-$T$ coherent electrons is found in heavy fermion compounds and many other strongly correlated materials.

Comparison with  $d-d$-excitons sheds considerable light on intraband excitons. 
In Fig.~\ref{fig:2}, we note that the largest excitonic gap arises when the two bands are exactly degenerate.  Hence, we should expect nearly identical excitonic physics in the single-band Hubbard model, where the two dispersions correspond to the spin-up and spin-down electrons.  But in this case, the significant role of the VHS is well known.  Figure~\ref{fig:12}(a) resembles a semiconductor with $E_g'=0$.  However, in this case it actually represents a single band with $E_F=E_{VHS}$, showing the VHS dispersions parallel to the $x$-axis (along the $y$-axis, the two dispersions are interchanged).

The close connection between the above $d-d$ excitons and intraband excitons can best be understood in terms of a Kugel-Khomskii-type model\cite{KuKho},
\begin{equation}
H=-\sum_{<i,j>}[J_s{\bf S_i\cdot S_j}+J_t{\bf T_i\cdot T_j}+4J_{st}({\bf S_i\cdot S_j})({\bf T_i\cdot T_j})]
-\sum_i[h_sS_i^z+h_tT_i^z],
\label{eq:5b}
\end{equation}
where ${\bf T}$ is a pseudospin operator representing orbital symmetry, with $T^z=+1/2$ corresponding to $d_{z^2}$ and $T^z=-1/2$ to $d_{x^2-y^2}$.  This Hamiltonian is formally symmetrical between the spin ${\bf S}$ and the pseudospin ${\bf T}$.  At half filling, when $h_t=0$, $h_s$ is large, we recover a [spinless] version of the $d-d$ exciton model, with orbital antiferromagnetic order, $T^z=1/2$ on one sublattice, -1/2 on the other.  For the opposite limit, $h_s=0$, $h_t$ large, we recover a single-orbital spin antiferromagnet, which must be equivalently excitonic, but intraband.

One should not expect Eq.~\ref{eq:5b} to hold exactly.  For instance, $h_s$ is a real magnetic field (or Hund exchange), while $h_t$ is a crystal field, and indeed the orbital bands in Fig.~\ref{fig:2} split into a total of four bands when orbital order is present.  However, in the special case of orbital degeneracy ($h_t\sim E_g=0$), Fig.~\ref{fig:12}(b), only two bands are present (darker shaded regions), exactly as in the one-band antiferromagnet problem.  Indeed, the actual $d-d$-excitonic dispersions in Fig.~\ref{fig:12}(b) bear a close resemblance to the antiferromagnetic cuprate bands (see Fig.~11(a) of Ref.~\onlinecite{AIP} and note that our choice of hopping parameters is appropriate for La$_2$CuO$_4$ with renormalization $Z=0.5$).  The only difference is that our choice of $V$ corresponds to a slightly larger value of $U$ than is appropriate for the cuprates; reducing the gap to $\Delta=0.5$~eV and shifting the Fermi level by -0.2~eV (light shaded regions) greatly enhances the similarity to cuprate experiment.  Note that if we replace $V_{k-k'}$ by the Hubbard $U$, Eq.~\ref{eq:5} becomes
\begin{equation}
1=U\chi_{Q},
\label{eq:5a}
\end{equation}
where $\chi_Q=\sum_{k'}\chi_{Q,k'}$, the usual Stoner criterion for a density-wave transition.  Thus, just as for the interband excitons, the mean-field exciton theory matches the corresponding density-wave theory.

We note that the vertex corrections in our pseudogap model are based on Moriya's mode-coupling model of quantum phase transitions.\cite{Moriya}  He introduced the model to better understand the transition between localized and itinerant spins, and by marrying the calculation to realistic band dispersions, we believe that we have taken a significant step towards that goal.

\subsection{A local model of the excitons}

While the $d-d$ and mode-coupling models capture many features of orbital and spin excitonic phases, they are mean field models with excitonic effects encoded in a renormalized susceptibility.  It would be more satisfying to be able to directly visualize the excitons.  Since 2D excitons should be very localized, a good place to look for them would be in a small sample,where there are only a few excitons present.  Here we study the smallest sample -- a $2\times 2$ plaquette -- for which a VHS can be defined.  Recent dynamic cluster approximation (DCA) calculations of the AFM state\cite{Jarr} offer support for this conjecture.   They find that the most accurate estimates of the ground state energy arise when the cluster contains an integer number of these $2\times 2$ plaquettes.

Hence we solve the mean-field Hubbard and $d-d$ models on a plaquette, focusing on $(\pi,\pi)$ instabilities [Appenoix].  For simplicity spin-orbit coupling is neglected.  We find that some of the electronic states are closely related to VHS nesting; we call these $V$-states.  Others seem to be more conventional fermionic states, acting mainly as spectators in VHS nesting, and we label these as $F$-states.  For the one-orbital Hubbard model, we find that the plaquette model captures many features of the susceptibility competition found in Ref.~\onlinecite{MBMB}.  Thus, in the Hubbard model, when $U$ is small the $F$-states can drive an AFM instability with small moments, as in a weak-coupling model, while when $U$ increases the ground state crosses over to the $V$-states and a large-moment AFM, thereby mimicing the Slater-Mott transition found in Ref.~\onlinecite{MBMB}.  The $V$-states are more localized (dispersion depends only on $t'$), and form excitons by coupling up-spins along one plaquette diagonal and down-spins along the other.  Moreover, when both $V$- and $F$-states are present, the latter develop a large moment, providing a clear picture of bosonic-dressed fermions.

When the model is extended to include $d_{z^2}$-electrons on equal footing with the $d_{x^2-y^2}$-electrons, the corresponding $V$-states continue to play a dominant, excitonic role.  In particular, when the two $d$-levels are degenerate, interorbital hopping of the $V$-states leads to a spin-ferromagnetic, antiferro-orbital ordering replacing the AFM order.  In contrast, electrons in the $F$-states have no direct coupling between the two orbitals.

At this point we can compare the $V$-states of the excitonic $d-d$ transition with those of the single-band Hubbard model, to better understand just what states constitute the excitons of the latter model.  In the $d-d$ model, the exciton involves a mixing of electrons in the conduction band ($a$ or $d_{x^2-y^2}$ states) with holes in the valence band ($b$ or $d_{z^2}$), represented by Eq.~\ref{eq:B6} and leading to a $(\pi,\pi)$ orbital AFM model.  In the Hubbard model, the $d_{z^2}$-orbital is replaced by an opposite-spin $d_{x^2-y^2}$ orbital, leading to the AFM order.  However, while the mean-field model captures the $a-b$ hybridization, via interatomic interorbital hopping\cite{CostSlat}, it fails to develop a coupling between the opposite-spin states, and that coupling arises indirectly, via the $F$-states, when they are occupied.  The origin of the coupling can be traced to the kinetic exchange $\tilde J\sim 4t^2/U$.

\section{Discussion}

\subsection{Localized spins as excitons}

For conventional excitons, there is a well-known variational model\cite{Kohn}.  Here we adapt this model to the one-band case, treating the down spins as the initially filled band.  We start with a single hydrogen molecule, using a real-space formalism,
\begin{equation}
|\Psi_{0,\downarrow}>=\prod_{i=A,B}c_{i,\downarrow}^{\dagger}|0>,
\label{eq:15b2}
\end{equation}
\begin{equation}
|\Psi_{1,\uparrow,\downarrow}>=\sum_{j=A,B}(u_jc_{j,\uparrow}^{\dagger}c_{j,\downarrow})|\Psi_{0,\downarrow}>\\
=(u_Ac_{A,\uparrow}^{\dagger}c_{B,\downarrow}^{\dagger}-u_Bc_{B,\uparrow}^{\dagger}c_{A,\downarrow}^{\dagger})|0>.
\label{eq:B11}
\end{equation}
where $A$ and $B$ label the two hydrogen atoms.  The first equality of Eq.~\ref{eq:B11} is in the electron-hole picture,  and can be considered as bound states of the electron and the respective correlation hole.  On the other hand, the second equality, in an electron-electron picture, has the Heitler-London form\cite{Fulde} when $u_A=u_B$.   Note that if the electrons differed in orbital character -- say $s$ vs $p$ -- rather than spin, this would be a model of tightly-bound Frenkel excitons.

If we reinterpret $A$ and $B$ in Eq.~\ref{eq:B11} as the diagonals of the elementary plaquette, this is a singlet model of the $V$ excitons.  Thus the intraband exciton would be an up-spin electron bound to a down-spin hole -- a viable model for a localized spin.  Note however, that the second form of Eq.~\ref{eq:B11} represents {\it a form of RVB\cite{RVB} wave function.}  However, it is distinct in being an RVB of the $V$-states only, and hence requires a distinct name.  Since the first form of  Eq.~\ref{eq:B11} identifies this state as an intraband exciton, we call it a resonating valence exciton (RVE).  This result is consistent with the QPGW model.\cite{AIP}  There we found a Landau-like quasiparticle Green's function $G_Z$, which suggested that only $Z\sim 0.5$ of the electrons contributed to electronic quasiparticles, and now we find that the remainder are tied up in RVEs.  It will be interesting to explore the connection between this intraband excitonic insulator and magnons\cite{PWA}.

In principle, one could follow the RVB scheme to tile the full lattice with RVEs, as a model of a strong excitonic insulator.  Here, however, we only note that a similar variational calculation yields the AFM mean field solution.  At half filling, we take our initial state as a full spin-down band, for which the Pauli exclusion forbids double occupancy, 
\begin{equation}
|\Psi_{0,\downarrow}>=\prod_kc_{k+Q,\downarrow}^{\dagger}c_{k,\downarrow}^{\dagger}|0>,
\label{eq:5b2}
\end{equation}
where $Q=(\pi,\pi)$ and $k$ is a vector in the $(\pi,\pi)$-magnetic Brillouin zone.  We then create electron-hole pairs with a down-spin hole and up spin electron in the same state:  
\begin{equation}
|\Psi_{Q,\uparrow,\downarrow}>=\prod_k(u_kc^{\dagger}_{k,\uparrow}c_{k,\downarrow}+v_kc_{k+Q,\uparrow}^{\dagger}c_{k+Q,\downarrow})|\Psi_{0,\downarrow}>\\
=\prod_k(u_kc_{k+Q,\downarrow}^{\dagger}c_{k,\uparrow}^{\dagger}+v_kc_{k+Q,\uparrow}^{\dagger}c_{k,\downarrow}^{\dagger})|0>.
\label{eq:B1}
\end{equation}
But this is just the excitonic model of Fig.~\ref{fig:2}, with $E_g=0$ and spin replacing orbital degeneracy.  Equation~\ref{eq:B1} was actually introduced as a variational model for antiferromagnetism, and interpreted in terms of an excitonic insulator, prior to the discovery of the cuprates\cite{VIK}.

\subsection{$V$-excitons and the VHS}
Finally, we can understand the intimate connection between the excitons and the VHS.  As noted above, an array of these $V$-excitons plaquettes can tile the full lattice, producing the $(\pi,\pi)$ AFM groundstate.  If the plaquettes were non-interacting, these states would lead to a $\delta$-function peak in the susceptibility, but due to residual interaction the peak is broadened to a logarithmic singularity.

These $V$-states share many properties of the VHS.  First, magnetic order near half filling starts at $U=0^+$, as expected for a diverging susceptibility.  Second, the $V$-states can support either FM or AFM order, consistent with VHS divergence both at $q=0$ (in density-of-states) or at $(\pi,\pi)$, with the latter dominant near half-filling.  While mathematically, these divergences of the VHS are well-known, the underlying physical reasons have never been clear. The $V$-states provide insight into the underlying molecular mechanisms.  Thus, the competition in the cuprates between VHS and conventional nesting\cite{MBMB} mirrors the competition between $V$- and $F$-states found here.  This is consistent with Ref.~\onlinecite{WHoSh}, who found that spin glass fluctuations are enhanced by going to a plaquette basis.

The close connection of the $V$-states to excitonic physics is very suggestive here.  For the excitonic insulator formed from the valence and conduction bands of a semiconductor, the resulting exciton involves a hybridization of the wave-functions of the two bands -- that is, the formation of a covalent bond.

Consider once more Mott's gedanken experiment of expanding a lattice of H-atoms.  While it has had a tremendous theoretical legacy, related experiments fail to verify his picture.  The problem is chemistry -- hydrogen forms molecules, so one never gets a lattice of H-atoms.  Attempts to do the reverse experiment, compressing a solid of $H_2$ molecules, also runs into problems.  The electrons start to delocalize, but by forming longer molecules.  The experiment is frustrated by the chemistry of covalent bond formation.

\subsection{Related calculations}
In the present paper, we provide a simple mean-field solution on a plaquette to illustrate the similarity to interorbital excitons of the $d-d$ model.  Of course, on a $2\times 2$ plaquette, the problem can be solved exactly, and the results have been used\cite{HKL,CDMFT0} to gain insight as to why cluster DMFT calculations work so well despite their limited $q$-resolution.  However, such calculations must be treated with caution, since the $2\times 2$ plaquette is known to be highly anomalous.  DCA calculations can track AFM order to considerably larger cluster size, finding the temperature dependence of the Neel correlation length from the size dependence of the effective Neel temperature.  While the data mostly fall on a smooth curve, the Neel temperature for the $2\times 2$ cluster is anomalously small, due to an accidental near-degeneracy with a state of valence bond order, which is absent in larger clusters.  Also, restriction to a $2\times 2$ cluster misses the effects of the $t''$ hopping parameter, important for describing the CDW order.\cite{MBMB}

Cluster extensions of DMFT (CDMFT) have developed a model for the pseudogap as a first-order transition between a correlated Fermi liquid and a non-Fermi liquid with a pseudogap (Ref.~\onlinecite{VCA} and references therein).  However, the role of the VHS was not clearly understood.  Our vertex corrected calculation\cite{MBMB} revealed the profound role of the VHS within the pseudogap phase, and a number of features of these results have been confirmed/extended by new CDMFT calculations.\cite{CDMFT1,CDMFT2}  In particular, both CDMFT calculations \cite{CDMFT1,CDMFT2} find that the pseudogap $T^*$ terminates at a doping $x^*$ close to the doping $x_{FS}$ at which the VHS crosses the Fermi level.  However, Ref.~\onlinecite{CDMFT2} defines a finite $T$ version $T_{FS}$ of the latter crossing, and shows that $T^*$ and $T_{FS}$ have very different doping dependencies away from the $T\rightarrow 0$ limit. In Ref.~\onlinecite{MBMB} it was shown that the two susceptibility divergences associated with the VHS split up at finite $T$ and have very different $T(x)$ dependencies.  In particular, the peak at $\Gamma$ has $T_{\Gamma}(x)\sim T_{FS}(x)$, while $T_{(\pi,\pi)}(x)\sim T^*$, due to Pauli unblocking, thereby evincing a much closer connection between the pseudogap and the VHS.  The CDMFT calculations find that for low $T$ and large $x$ a separation develops with $x^*<x_{FS}$; however, as noted below CDMFT averages over susceptibility, which can miss a weak transition.

The present approach to MBPT can provide insight into the CDMFT results on cuprates.  Since CDMFT averages the susceptibility over a substantial part of the Brillouin zone ($1/4$ for a $2\times 2$ plaquette), it can overlook susceptibility peaks that are restricted to a limited region of $q$-space, as found for many nesting instabilities.  Thus, in the correlated Fermi liquid regime Ref.~\onlinecite{MBMB} finds evidence for both incommensurate AFM order and a magnetic analog of the CDW order known in cuprates, both with nesting divergences in  restricted $q$-ranges, not reported in the CDMFT calculations.  In contrast, the VHS-related $(\pi,\pi)$ instability is spread over a broad near-$(\pi,\pi)$ plateau, which explains why traces of the instability persist after CDMFT averaging.

While Ref.~\onlinecite{CDMFT1} relates the pseudogap to the VHS, the earlier Ref.~\onlinecite{CDMFT0} by some of the same authors suggests a connection to excitonic physics.  Our analysis indicates that these are not competing explanations, but two different descriptions of the same underlying phenonenon.  The close connection of optimal pseudogap with nearly electron-hole symmetric states\cite{MBMB,CDMFT1,CDMFT2} points also toward excitonic physics.

The close similarity to a recent dynamic vertex approximation (D$\Gamma$A) calculation of the three-dimensional (3D) Hubbard model ($t'=0$)\cite{Kohn} should also be noted.  Thus, while both papers\cite{Kohn,MBMB} find a failure of conventional Moriya-Hertz-Millis theory to describe the results, Ref.~\onlinecite{MBMB} offers a simple correction.  The 3D Hubbard phase diagram $T(x)$ shows a transition from a commensurate $(\pi,\pi)$ order with large transition to an incommensurate magnetic phase with much lower transition temperature.  Near the transition, the correlation length has a local minimum, which was characterized as `not an indication of a decreasing correlation length', but an artifact arising from a two-peak structure in $\chi$.  While the weak minimum we had found for LSCO could be similarly characterized, tuning $t'$ revealed a flat-topped susceptibility and a complete collapse of $\xi\sim a$ (lattice constant).  We find the same collapse by tuning either $t'$ or $T$ at fixed $x$, which is as strong at finite $T$ as it is at $T=0$.  While Ref.~\onlinecite{Kohn} notes the role of Kohn points, a 3D version of double nesting,\cite{MLSB,Metz} it does not explain why the dominant $(\pi,\pi)$ nesting is not associated with a Kohn point.  In contrast, an excitonic origin via $(\pi,\pi)$ nesting should work equally well in 2D or 3D.

Finally, there is overlap between the present results and earlier strong coupling calculations based on $t-J$ or Heisenberg approximations to the Hubbard model.  These calculations also found an emergent order on the boundary of a commensurate $(\pi,\pi)$ AFM order, and an incommensurate $(\pi,\pi-\delta)$ order -- sometimes with $\delta =\pi$.  However, this latter order was a form of valence-bond order, and not an incommensurate AFM, consistent with the absence of Fermi surface nesting in the Heisenberg model.

\subsection{Comment on weak-coupling calculations}

Appendix E of Ref.~\onlinecite{CDMFT2} points out that weak-coupling calculations starting with an RPA susceptibility and adding first order or $G_0W_0$ self-energy corrections are unreliable because they produce peaks in the imaginary part of the self-energy at wrong energies. We had come to the same conclusion in our earlier work, and for this reason developed the self-consistent quasiparticle-GW (QPGW) model to correct this deficiency of the $G_0W_0$ scheme\cite{AIP}. Hence a more meaningful procedure would be to directly compare CDMFT and QPGW calculations, as in Ref.~\onlinecite{AIP}, see Figs. 55,56 and Fig. 25.    In addition to the self-consitent self-energy our DFT+MBPT model contains self-consistent vertex corrections which lead to a complete breakdown of the RPA assumption.\cite{MBMB}

The philosophies underlying the present mode-coupling and CDMFT schemes are quite different.  CDMFT stresses accuracy of local physics at the expense of $q$-resolution, and hence cannot capture fluctuation effects on length scales larger than the cluster size.  In contrast, our mode-coupling approach retains full $q$-dependence, so that we would expect differences from CDMFT when nesting or long-range fluctuations are important.  Mode-coupling should be similar to D$\Gamma$A calculations, which also stress accurate $q$-dependence and vertex corrections based on Moriyaesque $\lambda$ corrections\cite{RM70,Kohn}. For example, in 3D the D$\Gamma$A {\it critical exponents} are in good agreement with the exact exponents\cite{Metz} based on the double-nesting features introduced in Ref.~\onlinecite{MLSB}. Of course, the 2D case is more complex, since mode-coupling corrections are needed simply to satisfy the Mermin-Wagner theorem.\cite{MBMB}  Notably, a recent D$\Gamma$A calculation has demonstrated that the spurious Mott-Hubbard transition found in CDMFT arises from the inability of CDMFT to handle long-range correlations\cite{HT1} -- an apparent finite-$T$ Neel transition arises when the correlation length $\xi(T)$ is equal to the cluster size.\cite{Jarr}  Instead, D$\Gamma$A calculation\cite{HT1} finds that the gap opens when $U=0^+$, which is the VHS result for $t'=0$, fully compatible with mode-coupling physics.

Finally, the strength of correlations in the cuprates remains unclear. If the renormalization factor $Z$ is taken as a measure of this strength, as suggested by Brinkman and Rice\cite{BR}, then heavy fermions should be considered as being highly correlated with values of $Z\sim 0.1$, while the pnictides with $Z\sim 0.3$ would seem more correlated than the cuprates ($Z\sim 0.5$).  The cuprates, however, host the pseudogap, but if the pseudogap is driven by the VHS, then the underlying physics of the cuprates may not be that correlated.

\section{Summary and conclusions}

In conclusion, a number of recent calculations have produced a consistent picture of the cuprate phase diagram, both in 3D\cite{Kohn} and in 2D\cite{MBMB,CDMFT2,CDMFT1}, where an antiferromagnetic transition in 3D evolves into the 2D pseudogap.  All three 2D calculations find that the pseudogap terminates near the VHS, but that the intra-VHS scattering near $\Gamma$ has a distinct doping dependence from $T^*(x)$, while the doping dependence of the inter-VHS scattering near $(\pi,\pi)$ closely follows that of the pseudogap\cite{MBMB}.  Hence the key issue in pseudogap physics is to understand the origin of the anomalous inter-VHS scattering.  The present paper demonstrates the close connection between the inter-VHS scattering and excitonic physics, suggesting that the pseudogap is an excitonic insulator, and the pseudogap transition is related to the BEC-BCS transition.

Our purpose is to understand effects of realistic band structures on the excitonic insulator transition, and to ascertain if there is a single-band version of the excitonic transition that could be relevant to cuprate physics.  In this connection, even though we have considered a $d-d$ model in order to exploit the analogy between spin and orbital orders, our results are more general. Realistic band structures are shown to lead to significant modifications in excitonic transitions, revealing the key roles of level repulsion and inter-VHS coupling effects in driving these transitions.  Moreover, a clear connection between the traditional two-band excitonic transitions and a one-band excitonic model appropriate for the cuprates is identified.

The strong role of covalent bonding or hybridization suggests that first-principles band structure calculations could effectively capture many important features of the physics of correlated materials, even though the theory will be of limited reach in the strong coupling limit when excitons will be most localized. In this connection, here we have developed a microscopic model for the excitons in term of $V$-states, which are shown to resemble a particular form of RVB states.

Analogy between excitonic insulators and superconductivity might appear surprising because the treatment of excitonic insulators is traditionally based on a two-band model, whereas superconductivity is generally studied in a single-band model.  However, superconductivity can also exist in multi-band models, and an exact analog of the excitonic BEC has recently been discovered in a superconductor, referred to as `Fermi-surface-free superconductivity'\cite{Bang,Mist}, where the superconducting transition takes place even though one band is unoccupied in the normal state. Our study completes this analogy by developing an excitonic analog of single-band superconductivity.
\section*{Acknowledgements}     
This work is supported by the US Department of Energy, Office of Science, Basic Energy Sciences grant number DE-FG02-07ER46352, and benefited from Northeastern University's Advanced Scientific Computation Center (ASCC) 
and the allocation of supercomputer time at NERSC through grant number DE-AC02-05CH11231.  We thank Adrian Feiguin for stimulating discussions.  
\appendix
\section{Details of calculations}
\subsection{Plaquette Hubbard Model}

We begin by solving the mean-field Hubbard model on a single plaquette, assuming nearest neighbor $t$ and second neighbor $t'$ hopping,
\begin{equation}
  {\cal H} =
\left(
   \begin{array}{cccc}
\Delta_{1}& -t& -t'& -t\\
-t&\Delta_{2}&  -t & -t'\\
-t'&-t&\Delta_{3}&-t\\
-t& -t' &-t&  \Delta_{4}
   \end{array}
\right).
\label{eq:Ham} 
\end{equation}
We number the atomic sites as 1-4, starting in the lower-left corner and continuing clockwise, defining on-site orbitals $\psi_{i,\sigma}$, $i=1,4$ and $\sigma =\uparrow,\downarrow$.  Then, defining $\psi_{1\pm}=(\psi_1\pm\psi_3)/\sqrt{2}$, $\psi_{2\pm}=(\psi_2\pm\psi_4)/\sqrt{2}$, the eigenvalues [eigenfunctions] of the hopping Hamiltonian (all $\Delta$'s =0) are $E_{A,D}=-t'\mp 2t$ [$\psi_{A,D}=(\psi_{1+}\pm\psi_{2+})/\sqrt{2}$],
$E_B=E_C=t'$ [$\psi_{1-},\psi_{2-}$].  Since $E_B$ and $E_C$ are degenerate, we can also  write the eigenfunctions as the linear combinations $\psi_{B,C}=(\psi_{1-}\pm\psi_{2-})/\sqrt{2}$.   For this choice, the eigenvectors have the $k$-vector symmetry of the plaquette, $A=\Gamma$, $D=(\pi,\pi)$, $B=(\pi,0)$, and $C=(0,\pi)$.  For $U=0$, each level is filled with two electrons in order $E_A<E_B=E_C<E_D$.

Interaction causes these states to separate into two groups, which compete in a manner similar to the VHS- vs Fermi surface nesting competition found in cuprates.  Hence we label states $B$ and $C$ as $V$-states [VHS-related] since they fall at the VHS $k$-vectors $(\pi,0)$ and $(0,\pi)$ and correspond to the peak DOS, while states $A$ and $D$ are $F$-states [Fermi surface related] in that they depend on hopping $t$ as well as $t'$.
  
When the Hubbard $U$ is included at mean-field level, the energies are renormalized in a spin-dependent manner.  The lowest energy state corresponds to ${\bf Q}=(\pi,\pi)$ AFM order, and we assume that the excess up-spins are on atoms 1 and 3, $\Delta_i=\Delta_0+(-1)^i\Delta_m$, $\Delta_0=nU/2$, $\Delta_m=mU$, and
\begin{equation}
E_{k,\pm,\sigma}=\epsilon_++\Delta_0\pm\sqrt{\epsilon_-^2+\Delta_{m}^2},
\label{eq:C1}
\end{equation}
with $\epsilon_{\pm}=(\epsilon_k\pm\epsilon_{k+Q})/2$, $\sigma$ is the spin, $\bar\sigma$ the opposite spin, and $n_{\sigma}$ the density of spin $\sigma$ electrons on a given site.   The corresponding wavefunctions are 
$|\psi_{k\pm}|^2=(1\pm\Delta_m/\sqrt{\epsilon_-^2+\Delta_{m}^2})/2$.   
For ${\bf k}=(\pi,0)$, $U$ mixes it with ${\bf k}=(0,\pi)$, and since $\epsilon_-=0$, $U$ breaks the degeneracy and choses eigenfunctions $\psi_{1-}$ (for, e.g., up-spins) and $\psi_{2-}$ (for down-spins) independent of the magnitude of $U$.  Similarly, $U$ mixes $\Gamma$ and $(\pi,\pi)$, so on the up-spin sites $|\psi_{1+}|^2=(1+\Delta/\sqrt{4t^2+\Delta^2})/2$, $|\psi_{2+}|^2=(1-\Delta/\sqrt{4t^2+\Delta^2})/2$, with the signs of the second term reversed on the down-spin sites.  

\begin{figure}
\leavevmode
\epsfig{file=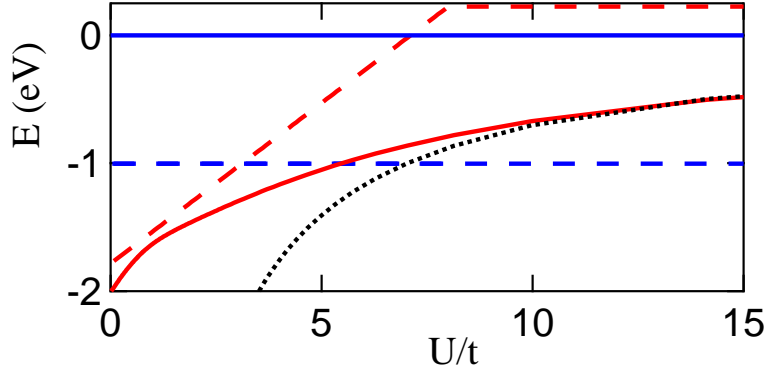,width=11.5cm, angle=0}
\vskip0.5cm  
\caption{ 
{\bf Plaquette energies for $N_0$ = 2 (dashed lines) and 4 (solid lines).}  Blue lines are for FM order, red for paramagnetic/AFM.  Black dotted line is -3.5J, with $J=4t^2/U$.
}
\label{fig:5}
\end{figure}

For $U\rightarrow\infty$ the energies are given by first filling the up-spin levels, in the same A-(B,C)-D order as above, then the down-spin levels.  While this yields ferromagnetic order away from half filling, there is no order at half filling, since states with different spins on each atom are degenerate.  Figure~\ref{fig:5} shows how the noninteracting levels (red lines) evolve with $U$, comparing them to the ferromagnetic levels (blue lines), as the number $N_0$ of electrons on the plaquette is varied, corresponding to electron doping $n=N_0/4$ per atom.  We focus on two cases: the dashed lines show $N_0$ = 2 (hole doping), and the solid lines $N_0=4$ (undoped). 

For $N_0=2$, the blue dashed line represents the $U$-independent ferromagnetic state, $E_A+E_B$, and the dashed red line the state that evolves from the spin-up and down $E_A$ states when $U$ is turned on.  Since both spins are occupied, $\Delta_0$ becomes finite, while $\Delta_m$ turns on when the magnetization $m$, given by
\begin{equation}
m=\frac{\Delta_mn/2}{\sqrt{4t^2+\Delta_m^2}},
\label{eq:C11}
\end{equation}
becomes finite.  Equation~\ref{eq:C11} can be rewritten as $m=\sqrt{n^2/4-4t^2/U^2}$, so $m$ is only finite for $U>8t$ -- this corresponds to the break of slope of the red line in Fig.~\ref{fig:5}.  Thus, for $N_0=2$, heavily hole-doped, the AFM only involves $F$-states, and is small-moment and metastable for all $U$.  There is a first order transition from the paramagnetic to the ferromagnetic state at $U\sim 3.1t$.

For $N_0=4$ the solid blue line corresponds to all four spin-up states occupied, while the red line is the corresponding AFM state.  In this state the spin-up $V$-state is on one diagonal sublattice, with the spin-down state on the other, and the two $A$ $F$-states are polarized as for $N_0=2$, but in the presence of the $V$-states.  This leads to an enhanced magnetization,
\begin{equation}
4m=1+\frac{\Delta_m}{\sqrt{4t^2+\Delta_m^2}},
\label{eq:C12}
\end{equation}  
where the first term on the right comes from  the $V$-states.  In this case the magnetization is strongly enhanced -- there is no longer a paramagnetic phase at finite $U$, and the AFM phase starts at $U=0$.  Moreover, the AFM phase is the ground state for all finite $U$.  The behavior of the $F$-states when the $V$-states are occupied can justly be described as a quasiparticle dressed by spin fluctuations.  Note that for large $U$, the difference betweem FM and AFM order is $\sim\tilde J =t^2/2\Delta_-\sim t^2/U$.

By confining the carriers to a 2$\times$2 plaquette, we have modified the effective band structure.  Thus, (1) hopping beyond 2d neighbors takes one off of the plaquette, and hence plays no role; (2) some first and second neighbor hopping also leads off of the plaquette, so the coefficients of the $t$ and $t'$ terms are anomalous.  This leads to an electron-hole symmetrical picture with the effective VHS ($\epsilon_{(\pi,0)}$) at the Fermi level at half filling.   However, the plaquette is not isolated but part of the larger crystal and in general the states $\psi_{1-}$, $\psi_{2-}$ are not at the Fermi level.  This suggests that these VHS-excitons are Pauli blocked at $T=0$ but can readily form at $T>T_{VHS}\sim E_F-E_{VHS}$.  This is quite similar to the phenomena we find in cuprates when mode-coupling is included.

\subsection{Plaquette $d-d$ Model}

The bare dispersion of Fig.~\ref{fig:4} is a single layer version of the model of Ref.~\onlinecite{Baubl},
\begin{equation}
{\cal H}=\Bigl({\begin{array}{cc}
                 H_a+E_z/2 & H_c \\
                 H_c & H_b-E_z/2
                 \end{array}}\Bigr),
\label{eq:21b}
\end{equation}
with $H_a=-2t_a(c_x(a)+c_y(a))-4t_a'c_x(a)c_y(a)-2t_a''(c_x(2a)+c_y(2a))-2t_a'''(c_x(3a)+c_y(3a))$, $H_b=-2t_b(c_x(a)+c_y(a))-2t_b'c_x(a)c_y(a)$, $H_c=-2t_c(c_x(a)-c_y(a))-2t_c''(c_x(2a)-c_y(2a))-2t_c'''(c_x(2a)c_y(a)-c_y(2a)c_x(a))$ (compare Eq.~\ref{eq:1b}).  The hopping parameters are [in meV] $(t_a,t_a',t_a'',t_a''') =  (502,-112,92,21)$, $(t_b,t_b')=(170,75)$, and $(t_c,t_c'',t_c''')=(251,13,14)$, while $E_z$ is the gap parameter $E_g$ of Fig.~\ref{fig:4} -- in the bilayer manganates, $E_z=-305$~meV.

In Eq.~\ref{eq:Ham}, we introduced a simple plaquette model for the intraband excitons.  Here we extend the results to two $d$-orbitals, to form the plaquette model appropriate for Fig.~\ref{fig:4}, but explicitly including spin.  In this case, the Hamiltonian becomes an $8\times 8$ matrix:
\begin{equation}
{\cal H}=\Bigl({\begin{array}{cc}{\cal A}&{\cal B}\\
{\cal B}&  {\cal A}_{z}
                 \end{array}}\Bigr).
\label{eq:B4}
\end{equation}
 In this equation, ${\cal A}$, ${\cal A}_z$, and ${\cal B}$ are $4\times 4$ matrices of the same form as that in Eq.~\ref{eq:Ham}.  ${\cal A}$ represents the $d_{x^2-y^2}$ orbitals, and is identical to Eq.~\ref{eq:Ham} with hopping parameters $t_a$, $t_a'$ (below Eq.~\ref{eq:21b}), whereas ${\cal A}_z$ represents the $d_{z^2}$ orbitals, with hopping $t_b$, $t_b'$, and a $d_{x^2-y^2}-d_{z^2}$-splitting $E_z$ added to the diagonal terms.  The mixing term ${\cal B}$ is 
\begin{equation}
  {\cal B} =
\left(
   \begin{array}{cccc}
0& t_c& 0& -t_c\\
t_c&0&  -t_c& 0\\
0&-t_c&0&t_c\\
-t_c& 0 &t_c&  0
   \end{array}
\right).
\label{eq:B5} 
\end{equation}   
In the plaquette calculations, we label the atomic sites by a number, 1-4, as in the above Subsection, and the orbitals by a letter $a$ for $d_{x^2-y^2}$ or $b$ for $d_{z^2}$.  Just as in the $4\times 4$ calculation, the present Hamiltonian can be simplified by forming wavefunctions $\psi_{1\pm}$, etc., thereby reducing the $8\times 8$ to four $2\times 2$ matrices.  One of these matrices is the F-states of ${\cal A}$, identical to those of Eq.~\ref{eq:Ham}, and a second one the F-states of ${\cal A}_z$, differing only in the parameter values.  

The other two matrices are much more interesting: the ${\cal B}$-matrix mixes the V-states of the two orbital bands, producing the interband VHS mixing revealed in Figs.~\ref{fig:2}-\ref{fig:4}.  We discuss the mixing of $\psi_{a1-}$ with $\psi_{b2-}$ (labeled $\alpha$), but the mixing of $\psi_{a2-}$ with $\psi_{b1-}$ ($\beta$) is analogous.  Defining $\tilde\Delta_{\alpha\pm}=(\Delta_{a1}\pm\Delta_{b2})/2+t'_{\pm}$, $t'_{\pm}=(t_a'\pm t_b')/2$, the eigenvalues are $E_{\alpha\pm}=\tilde\Delta_{\alpha +}\pm\sqrt{\tilde\Delta_{\alpha -}^2+4t_c^2}$, with eigenfunctions
\begin{equation}
\begin{array}{c}
\psi_{\alpha -}=u_{\alpha -}\psi_{a1-}+v_{\alpha -}\psi_{b2-}
\\
\psi_{\alpha +}=v_{\alpha -}\psi_{a1-}-u_{\alpha -}\psi_{b2-},
\end{array}
\label{eq:B6}
\end{equation}
with $u_{\alpha -}^2=1-v_{\alpha -}^2=(1+\tilde\Delta_{\alpha -}/\sqrt{\tilde\Delta_{\alpha -}^2+4t_c^2})/2$.  Note that this is exactly the excitonic form of Eq.~\ref{eq:B1}.  As $E_z$ varies, we can follow the same evolution of the bands, but restricted to the $V$-states.  In particular, the special case $E_z+t'_-=0$ corresponds to degenerate $a$ and $b$ orbitals, as in Fig.~\ref{fig:4}; from  Eq.~\ref{eq:B6}, $u_{\alpha -}^2=1-v_{\alpha -}^2=(1-(E_z+t'_-)/\sqrt{(E_z+t_-)^{'2}+4t_c^2})/2$=1/2.   Thus, the physics of the $V$-states can properly capture orbital antiferromagnetic (plus spin ferromagnetic) ordering.


\begin{thebibliography}{99}



\bibitem{McMi}W.L. McMillan, 
{ Phys. Rev. B}{\bf 16,} 643 (1977).
\bibitem{Moti}K. Motizuki and N. Suzuki,  
{\it Structural Phase Transitions in Layered Transition-Metal Compounds} (Reidel, Dordrecht, 1986).
\bibitem{Moriya}T. Moriya, 
{\it Spin Fluctuations in Itinerant Electron Magnetism}
(Springer, Berlin, 1985).
\bibitem{MBMB}R.S. Markiewicz, I.G. Buda, P. Mistark, and A.  Bansil, 
{ Nature Scientific Reports} {\bf 7,} 44008 (2017) (arXiv:1505.4770).
\bibitem{Mott}N.F. Mott,  
{ Phil. Mag.} {\bf 6,} 287 (1961). 
\bibitem{KeKu} L.V. Keldysh and Yu.V. Kopaev,  
{ Sov. Phys. Sol. State} {\bf 6,} 2219 (1965). 
\bibitem{Kohn} W. Kohn, 
in {\it Many Body Physics}, edited by C. DeWitt and R. Balian (Gordon \& Breach, New York, 1968), p. 351.
\bibitem{HaRi}B.I. Halperin and T.M. Rice, 
in {\it Solid State Physics}, Vol.~21, ed. F. Seitz, D. Turnbull, and H. Ehrenreich 
(New York, Academic) p. 115.
\bibitem{CoG}R. C\^ot\'e and A. Griffin, 
{ Phys. Rev. B}{\bf 37,} 4539 (1988).
\bibitem{BroF}F.X. Bronold and H. Fehske, 
{ Phys. Rev. B}{\bf 74,} 165107 (2006).
\bibitem{Kohn2}W. Kohn, 
{ Phys. Rev. Lett.} {\bf 19,} 789 (1967).
\bibitem{Mott2}N.F. Mott, 
{ Rev. Mod. Phys.} {\bf 40,} 677 (1968).
\bibitem{JCP}J.C. Phillips, 
{ Phys. Rev.} {\bf 136,} A1705 (1964).
\bibitem {Rief}A. Riefer, F. Fuchs, C. R\"odl, A. Schleife, F. Bechstedt, and R. Goldhahn,  
{ Phys. Rev. B}{\bf  84,} 075218 (2011).
\bibitem{Baubl}M. Baublitz, C. Lane, Hsin Lin, Hasnain Hafiz, R.S. Markiewicz, B. Barbiellini, Z. Sun, D.S. Dessau, and A. Bansil, 
{ Nature Sci. Rep.} {\bf 4,} 7512 (2014). 
\bibitem{Paco}J.A. Silva-Guill\'en, P. Ordej\'on, F. Guinea, and E. Canadell, 
2D Materials {\bf 3}, 035028 (2016). 



\bibitem{AIP}T. Das, R.S. Markiewicz, and A. Bansil,  
{ Advances in Physics} {\bf 63,} 151 (2014).
\bibitem{VHex1} R.S. Markiewicz,  
{ Physica C} {\bf 168,} 195 (1990);
{ J. Phys.: Cond. Matt.} {\bf 3,} 3859 (1991).
\bibitem{VHex2}F. Onufrieva and P. Pfeuty, 
{ Phys. Rev.}{\bf B 61,} 799 (2000).


\bibitem{KuKho}J. van den Brink, W. Stekelenburg, D.I. Khomskii, G.A. Sawatzki, and K.I.  Kugel, 
{ Phys. Rev. B}{\bf 58,} 10276 (1998).




\bibitem{Jarr}T.A. Maier, M. Jarrell, T.C. Schulthess, P.R.C. Kent, and J.B. White,  
{ Phys. Rev. Lett.} {\bf 95,} 237001 (2005).
\bibitem{CostSlat}J.C. Slater and G.F. Koster,  
{ Phys. Rev.} {\bf 94,} 1498 (1954).


\bibitem{Fulde}P. Fulde, 
{\it Electron Correlations in Molecules and Solids}, 2d Edition
(Springer, Berlin, 1993).
\bibitem{RVB}P.W. Anderson, 
{ Science} {\bf 235,} 1196 (1987). 
\bibitem{PWA}C.-M. Ho, V.N. Muthukumar, M. Ogata, and P.W. Anderson,  
{ Phys. Rev. Lett.} {\bf 86,} 1626 (2001).
\bibitem{VIK}S.V. Vonsovsky, V.Yu. Irkhin, and M.I. Katsnelson, 
{ J. Mag. Mag. Mats.} {\bf 58,} 309 (1986). 

\bibitem{WHoSh}P. Werner, S. Hoshino, and S. Shinaoka, 
Phys. Rev. B{\bf 94}, 245134 (2016). 
\bibitem{HKL}M. Harland, M.I. Katsnelson, and A.I. Lichtenstein, 
{ Phys. Rev. B}{\bf 94,} 125133 (2016).
\bibitem{CDMFT0}S. Sakai, M. Civelli, and M. Imada, 
{  Phys. Rev. B}{\bf 94,} 115130 (2016).
\bibitem{VCA}M. Balzer, B. Kyung, D. S\'en\'echal, A.-M.S. Tremblay, and M. Potthoff, 
{ EuroPhys. Lett.} {\bf 85,} 17002 (2009).
\bibitem{CDMFT1}H. Braganca, S. Sakai, M.C.O. Aguiar, and M. Civelli, 
to be published, Phys. Rev. Lett. 
\bibitem{CDMFT2}W. Wu, M.S. Scheurer, S. Chatterjee, S. Sachdev, A. Georges, and M. Ferrero, 
{ arXiv:1707.06602}.
\bibitem{Kohn}T. Sch\"afer, A.A. Katanin, K. Held, and A. Toschi, 
{ Phys. Rev. Lett.} {\bf 119,} 046402 (2017). 
\bibitem{RM70}R.S. Markiewicz, 
{ Phys. Rev. B}{\bf 70,} 174518 (2004).
\bibitem{Metz}T. Holder and W. Metzner, 
Phys. Rev. B {\bf 90}, 161106(R) (2014).
\bibitem{MLSB}R.S. Markiewicz, J. Lorenzana, G. Seibold, and A. Bansil,  
{ Phys. Rev. B}{\bf 81,} 014509 (2010).
\bibitem{HT1}T. Sch\"afer, F. Geles, D. Rost, G. Rohringer, E. Arrigoni, K. Held, N. Bl\"umer, M. Aichhorn, and A. Toschi, 
{ Phys. Rev. B}{\bf 91,} 125109 (2015).
\bibitem{BR}W.F. Brinkman and T.M.Rice, 
{ Phys. Rev. B}{\bf 2,} 4302 (1970).

\bibitem{Bang}Y. Bang,  
 New J. Phys. {\bf 16}, 023029 (2014). 
\bibitem{Mist}P. Mistark, H. Hafiz, R.S. Markiewicz, and A. Bansil,  
{ Nature Scientific Reports} {\bf 5,} 9739 (2015).

\end{thebibliography}
\end{document}